\documentclass{ifacconf}

\usepackage{graphicx}      
\usepackage{natbib}        

\usepackage{amsmath}
\usepackage{amssymb}
\usepackage{algorithm,algorithmic}
\usepackage[center]{subfigure}
\usepackage{euscript}
\usepackage{verbatim}
\usepackage{amssymb}
\usepackage{color}
\usepackage{cancel}
\usepackage{calligra}
\usepackage{pgf}
\usepackage{tikz}
\usepackage{ctable}
\usetikzlibrary{arrows,automata}
\usetikzlibrary{shapes}
\usetikzlibrary{positioning}
\usepackage{bbm}
\usepackage{multirow}
\usepackage{url}

{\normalfont}{\normalfont}
{\normalfont}{\normalfont}
{\normalfont}{\normalfont}

\newcommand{\mycite}[1]{\citeauthor{#1}, \citeyear{#1}}

\newcommand{\arnew}[1]{\textcolor{blue}{#1}}

\begin{document}

\begin{frontmatter}

\title{SINDy vs Hard Nonlinearities and Hidden Dynamics: a Benchmarking Study}


\author[First]{Aurelio Raffa Ugolini} 
\author[Second]{Valentina Breschi} 
\author[Third]{Andrea Manzoni}
\author[First]{Mara Tanelli}

\address[First]{Dip. di Elettronica, Informazione e Bioingegneria, Politecnico di Milano, Via G. Ponzio 34/5 - 20133 Milano, Italy.}
\address[Second]{Dept. of Electrical Engineering, Eindhoven University of Technology, 5600 MB Eindhoven, The Netherlands.}
\address[Third]{Dip. di Matematica, Politecnico di Milano, P.zza Leonardo da Vinci 32- 20133 Milano, Italy.}

\begin{abstract}                
In this work we analyze the effectiveness of the Sparse Identification of Nonlinear Dynamics (SINDy) technique on three benchmark datasets for nonlinear identification, to provide a better understanding of its suitability when tackling real dynamical systems. While SINDy can be an appealing strategy for pursuing physics-based learning, our analysis highlights
difficulties in dealing with unobserved states and non-smooth dynamics.
Due to the ubiquity of these features in real systems in general, and control applications in particular, we complement our analysis with hands-on approaches to tackle these issues in order to exploit SINDy also in these challenging contexts.
\end{abstract}


\end{frontmatter}
\section{Introduction}

Model discovery in the time domain is challenging and often demands substantial prior knowledge of the system under consideration. Nonetheless, the growing availability of data and computational power has led to a shift towards data-driven methods, often at the expense of interpretability (see \mycite{Rudin2019}). Among existing approaches, the \emph{Sparse Identification of Nonlinear Dynamics} (SINDy, \mycite{Brunton2016}) promises to merge the benefits of data-driven techniques with (partial) expert knowledge, allowing to identify sparse nonlinear models via modest amounts of data, while incorporating priors on the targeted systems. Notably, this algorithm results in a closed-form expression of the dynamical model, thus making its outputs intelligible. For this reason among others, SINDy has catalyzed the attention of researchers ever since its first appearance, yielding several extensions and generalizations (see e.g., \mycite{2022faselEnsembleSINDyRobustSparse}; \mycite{2020kahemanSINDyPIRobustAlgorithm}; \mycite{2021messengerWeakSINDyGalerkinBased}; \mycite{championDatadrivenDiscoveryCoordinates2019}).\\
Although SINDy was originally introduced for the identification of autonomous systems, in \mycite{Brunton2016b} it has been extended to incorporate control actions\footnote{In the following we use the acronym SINDy to refer to the version with control inputs for the sake of brevity.}, making it another tool at the disposal of designers to learn models to be later used for control. Starting from this work, the effectiveness of SINDy in learning for (model predictive) control has indeed been evaluated in several works (see \mycite{Kaiser2018}; \mycite{Fasel2021}; \mycite{Schmitthenner2021}; \mycite{Lore2023}), considering applications ranging from flight to epidemic control. However, these studies often focus on noise-free scenarios (see \mycite{Fasel2021}, Section IV), lacking conclusiveness towards real-world control challenges.

With this work, we aim to provide further insights into the advantages and drawbacks of SINDy for system identification using three benchmark datasets, featuring (i) unobservable states, (ii) non-smooth dynamics/hard nonlinearities.
Specifically, we carry out our analyses on the Pick and Place machine dataset from \cite{Juloski2004}, the Bouc-Wen model from~\cite{2020noelHystereticBenchmarkDynamic} and Cascaded Tanks data from~\cite{cascadedTanksBenchmark}.
Each of them poses unique challenges and requires different hands-on strategies to address the limitations of the SINDy algorithm. This choice ultimately distinguishes our work from others, since we test SINDy on benchmark data specifically collected for the identification of nonlinear, controlled systems (see \emph{e.g.,} \mycite{MASTI2021109666}; \mycite{Bemporad2018}; \mycite{FORGIONE202169}).\\
The aim is to guide practitioners and researchers on what to expect when using SINDy, highlighting common challenges and potential research directions. The key insight is that SINDy can be a potent modeling tool, if users are aware of its common pitfalls, and steer away from the idea that it is a universally applicable tool.

The paper is organized as follows. An introductory description of SINDy and its main properties is provided in Section~\ref{sec:intro_sindy}, laying the technical grounds for our analysis. Section~\ref{sec:sindy-in-practice} introduces the strengths and weaknesses of SINDy, which are further discussed in Sections~\ref{sec:pick-and-place}-\ref{sec:cascaded-tanks}. These sections discuss our findings for each benchmark dataset, presented by increasing levels of insight into the actual dynamics we found necessary to achieve satisfactory results with SINDy. Based on these findings, in Section~\ref{sec:open-challenges} we draw concluding remarks and discuss future research directions.

\section{A brief overview of SINDy}\label{sec:intro_sindy}
Consider the following class of continuous-time systems
\begin{equation}\label{eq:system}
	\dot{x}(t)=f(x(t),u(t)),~~x(0)=x_{0},
\end{equation} 
where $x(t) \in \mathbb{R}^{n}$ is the system's internal state, $u(t) \in \mathbb{R}^{m}$ is an external, controllable input that can is fed to the system and $f:\mathbb{R}^{n} \times \mathbb{R}^{m}\rightarrow \mathbb{R}^{n}$ is the (unknown) function dictating the system dynamics. SINDy aims to identify a model for the system~\eqref{eq:system} from state and input measurements, exploiting a \textit{library} $\phi(x(t),u(t)) \in \mathbb{R}^{n_\phi \times 1}$ of \emph{predetermined} basis functions to yield an approximation of the unknown relationship \eqref{eq:system} in the form:
\begin{equation}\label{eq:model}
	\dot{x}(t) \approxeq \Theta^{\top}\phi(x(t),u(t)),
\end{equation}
where $\Theta \in \mathbb{R}^{n_{\phi} \times n}$ is the \textit{sparse} matrix including the coefficients associated with each of the basis functions in the user-defined library, which can be either linear or (smooth) nonlinear in the state, inputs or both of them.\\ 
Given a dataset $\{x_{k},x_{k}^{d},u_{k}\}_{k=1}^{T}$, where $x_{k}$, $x_{k}^{d}$ and $u_{k}$ denote the available samples of the state, its derivative and input trajectory, let us introduce the matrices
$$\dot{X}\!=\!\begin{bmatrix}x_{1}^{d} & x_{2}^{d} & \cdots & x_{T}^{d}\end{bmatrix}^{\top},\quad  X\!=\!\begin{bmatrix}x_{1} & x_{2} & \cdots & x_{T}\end{bmatrix}^{\top},$$
$$U\!=\!\begin{bmatrix}u_{1} & u_{2} & \cdots & u_{T}\end{bmatrix}^{\top},\quad\Theta=\begin{bmatrix}
		\Theta_{1} & \Theta_{2} & \cdots &\Theta_{n}
	\end{bmatrix}.$$
The problem of learning a sparse representation of \eqref{eq:system} with \eqref{eq:model} tackled by the SINDy algorithm amounts to the following sparse optimization problem:
\begin{equation}\label{eq:fitting_problem}
	\min_{\Theta}~~\big\|\dot{X}-\Phi(X,U)\Theta\big\|_{2}^{2}+\lambda^2 \sum_{i=1}^{n}\|\Theta_{i}\|_{0},
\end{equation}
where $\Phi(X,U)$ stacks (row-wise) the values of $\phi(x(t),u(t))$ computed for each input/state pair. Note that this problem is separable for the columns of $\Theta$ and, thus, its solution can be solved by tackling the resulting $n$ (sparse) regression problems in parallel. By looking at the fitting problem in \eqref{eq:fitting_problem}, it is clear that SINDy requires to $(i)$ estimate the state derivative, and $(ii)$ select the library of basis functions. While the state derivative can be computed with existing approaches (\emph{e.g.,} filtering), the choice of appropriate basis function heavily depends on the users' domain knowledge. As such, when this information is not available or not accurate enough, lengthy trial-and-error procedures are required to select a \textquotedblleft good\textquotedblright \ set of basis functions. Notably, \eqref{eq:fitting_problem} features a $\ell_{0}$-norm regularization, making 
the overall problem non-convex. A possible approach to handle this class of problems (introduced in the seminal work \mycite{Brunton2016b}) is the \emph{Sequential Thresholded Least Squares} (STLS) algorithm, whose iterative steps can be summarized as:
\begin{subequations}\label{eq:stls}
	\begin{align}
		& S^{j} \leftarrow \left\{h \in [1,n_{\phi}]: \big|\Theta_{h}^{(j)}\big|\geq\lambda\right\},\\
	& \Theta_{h}^{(j+1)} \leftarrow \arg\!\min_{\Theta_{h} \in S^{(j)}} \frac{1}{2}\big\|\dot{X}_{h}-\Phi(X,U)\Theta_{h}\big\|_{2}^{2},
	\end{align}
\end{subequations}       
for each column of $\Theta$ and, thus, $h=1,\ldots,n$. When $\Phi(X,U)$ is full rank, the scheme in \eqref{eq:stls} is known to converge to a fixed point in a finite number of steps that is upper-bounded by the dimension $n_{\phi}$ of the library, that is also a local minimizer of \eqref{eq:fitting_problem} (the reader is referred to \mycite{Zhang2019} for more details).   

To the best of our knowledge, said method has no theoretical guarantees of convergence to the true system structure and parameters, even when the actual bases of the system are included in the library. Convergence guarantees are instead provided for \emph{weak-SINDy} (\mycite{Russo2022}), an integral formulation of SINDy, under the somewhat limiting assumption that the true system is scalar and only polynomial and/or harmonic basis functions compose the library. However, the price to pay for having these guarantees is a more involved definition of the library and a less scalable optimization problem to solve.           

\section{A practical perspective on SINDy}\label{sec:sindy-in-practice}

Unlike many identification approaches that yield discrete-time models, SINDy facilitates the direct identification of continuous-time systems from sampled data. It supports nonlinear continuous-time identification, aiming to recover the presumed \textquotedblleft true\textquotedblright model structure by providing closed-form differential expressions, which can be sparsified through the introduced regularization term in~\eqref{eq:fitting_problem}.\\ 
Despite these potential advantages, successful model recovery, even under the sparsity assumption, hinges on both the library and coordinate selection, requiring expertise in the tuning process.
Moreover, SINDy encounters challenges in handling hard nonlinearities and unobserved states, prevalent in real-world (control) applications.  Indeed, hard nonlinearities are generally dictated by the physical limits of the system and the actuators, and (possibly) by the scale of the sensors, while accessing the full state of a system might be economically or practically infeasible.\\
Since conventional strategies to cope with hidden states (e.g., using Kalman filtering as in \mycite{MOHITE2021102152}) are not applicable in this context, extensions such as in~\cite{2022bakarjiDiscoveringGoverningEquations} have been proposed. This approach relies on deep autoencoders to craft a new set of states from time-delay embeddings of the observed outputs. However, by using a black-box model to overcome the limitations of SINDy, the method falls short of one of SINDy's core benefits, i.e., returning interpretable models.\\
Alternative strategies do exist to reconstruct hidden states by appropriately manipulating input/output data, but they demand $(i)$ deep system knowledge, $(ii)$ low-noise conditions (see \mycite{8836606}), or $(iii)$ significant trust in first principle models, restricting SINDy's practical applicability.

\section{Procedures and evaluation metrics}\label{sec:case_studies}
In our tests, we employ the open-source package PySINDy (\mycite{Pysindy2}). In each example, we select the most relevant hyperparameters via Bayesian optimization using the Hyperopt Python package. The final performance is then evaluated over a separate test set, as provided by each benchmark suite.\\
Note that the SINDy method requires the numerical approximation of the signals' derivatives. This operation can be critical if significant levels of noise corrupt the data, not allowing practitioners to successfully employ simple numerical approaches (e.g., finite differences). Although none of the examples features extreme levels of noise, to overcome this issue we add a tunable regularization of the derivatives that is optimized jointly with the other model hyperparameters\footnote{The code is available at \url{https://github.com/aurelio-raffa/benchmarking_sindy}.}. The performance of the SINDy approach is quantitatively assessed by 
one or both of the following quality indicators:
\begin{subequations}
	\begin{equation}
		\mbox{BFR}=\max\left(1-\frac{\sum_{k=1}^{T^{\mathrm{test}}}(y_{k}-\hat{y})^2}{\sum_{k=1}^{T^{\mathrm{test}}}(y_{k}-\bar{y})^2},0\right),~~[\%]
  \end{equation}
  \begin{equation}
		\mbox{RMSE}=\sqrt{\frac{1}{T^{\mathrm{test}}}\sum_{k=1}^{T^{\mathrm{test}}}(y_{k}-\hat{y}_{k})^2},
	\end{equation}
\end{subequations}
where $T^{\mathrm{test}}$ is the length of the test set, $\bar{y}$ is the mean value of the test output, and $y_{k}$, $\hat{y}_{k}$ are the actual test output and the one obtained in open-loop simulation with the fitted model, respectively, for $k=1,\ldots,T^{\mathrm{test}}$. The choice of the specific performance metric is based on the characteristics of the considered benchmark.

\section{Derivative-coordinate state representations: the Pick-and-place case}\label{sec:pick-and-place} 
We start our analysis by considering the dataset introduced in \cite{Juloski2004}, featuring input/output data collected over an experiment of 15~[s] (at a sampling frequency of $400$~Hz) from a pick-and-place machine in the act of placing an electronic component on a circuit board and then releasing it. The experimental apparatus is composed of a mounting head free to move vertically, actuated by an electric motor (whose input voltage represents $u(t)$ [V]). The position $y(t)$ of the head is measured over time, constituting the output. In this system, four operating modes can be detected: $(i)$ an \emph{upper saturation mode}, corresponding to the head being fully retracted; $(ii)$ a \emph{free mode}, namely with the head neither being retracted nor in contact with the impacting surface; $(iii)$ an \emph{impact mode}, when the head is in contact with the circuit board, but not at a saturation level yet; $(iv)$ a \emph{lower saturation mode}, \emph{i.e.,} the head is fully extended.\vspace{-.5cm}

\paragraph*{\textbf{Challenges and goal.}}
In the two saturation modes, the system's dynamics change abruptly, leading to a difficult modeling scenario for SINDy. Therefore, our main goal is to assess if the nonlinear modeling capabilities of SINDy are sufficient to compete with mode-switching algorithms, often assessed on this benchmark due to the nature of the underlying system (see, e.g., \mycite{FERRARITRECATE2003205}; \mycite{Bemporad2018}). This benchmark case study is also intended to show how mild priors on the system affect the accuracy of the learned model.\vspace{-.5cm}

\paragraph*{\textbf{Hands-on strategy.}} The system's dynamics, see \cite{Juloski2004}, describes the mounting head acceleration whenever the output is not in a saturated mode. We can thus assess the potential advantages of increasing the model order to provide a better approximation of the underlying dynamics. In addition to the output (the only state of the model in \eqref{eq:model_pick1}), we introduce the additional state
\begin{equation}\label{eq:true}
	z(t)=\dot{y}(t),
\end{equation}
whose measurements are artificially constructed from the available data by derivation. The full model combines~\eqref{eq:true} and the new state equation
\begin{equation}\label{eq:model_pick2}
\dot{z}(t)=\ddot{y}(t)=\phi^{\mathrm{pick}}(y(t),z(t),u(t))^{\top}\Theta_{z}. \end{equation} 
This approach is inspired by the well-known equivalence between a linear system of $n$ ordinary differential equations (ODE) and a single ODE of order $n$ already exploited in~\cite{2022somacalUncoveringDifferentialEquationsb}. \vspace{-.2cm}

\begin{figure}[!tb]
	\centering
    \includegraphics[scale=.475]{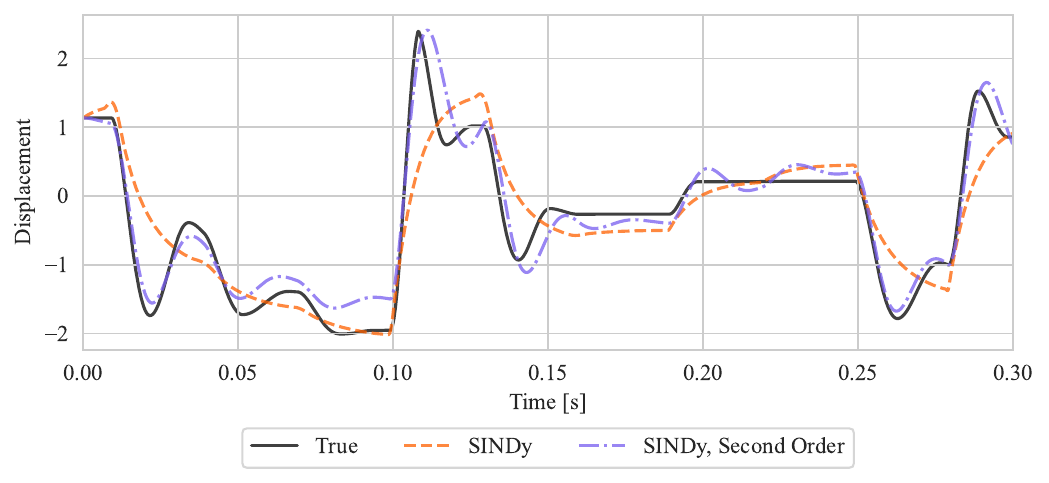}\vspace{-.3cm}
	\caption{Pick-and-place: simulation over the test set.}\label{fig:pick}
\end{figure}

\paragraph*{\textbf{Results.}}
By considering the standard SINDy library of $2$-nd order polynomial functions, we split the available data into a training set of length $3840$ samples, a validation set of $960$ data points and a test set of $1200$ samples (as done in \mycite{Bemporad2018}). The model returned by SINDy with these settings is 
\begin{equation}\label{eq:model_pick1}
	\dot{y}(t) = -111.16\ y(t) + 133.73\ u(t) - 3.48\ y(t) u(t).
\end{equation} 
As shown in \figurename{~\ref{fig:pick}}, the model results in an output that features slower dynamics with no oscillations and no plateaux. At the same time, its performance is comparable with those attained by two competitor approaches, namely the methods proposed in \cite{FERRARITRECATE2003205} to learn piecewise affine (PWA) models and the one presented in \cite{Bemporad2018} to fit jump models from data, respectively.\\
Instead, with the proposed hands-on strategy, using an identical library we obtain the behavior displayed in purple in \figurename{~\ref{fig:pick}}. The new model captures the output oscillations and better matches its behavior under saturation. Accordingly, this second-order model outperforms both \eqref{eq:model_pick1} and the competitor approaches, despite the hard nonlinearities, at the price of an increase in the model complexity.\vspace{-.5cm}

\paragraph*{\textbf{Take-home message.}} From the obtained results we can conclude that SINDy can recover good approximations of the dynamics when the system of equations can be expressed in terms of finitely many derivatives of the recorded output, as is often the case for mechanical systems, see also~\cite{2022somacalUncoveringDifferentialEquationsb}. At the same time, relying on higher-order derivatives of the state poses new challenges since $(i)$ it amplifies the effects of measurement noise and, more importantly, $(ii)$ state derivatives have to be estimated at inference time. These limitations can eventually be addressed via filtering, computing regularized derivatives, or employing weak formulations.

\begin{table}[!tb]
	\caption{Pick-and-place: BFR [\%] in testing.}\label{tab:pick}
	\centering
    \vspace{-.1cm}
	\begin{tabular}{lc}
        \textbf{Model} & \textbf{BFR [\%]}\\
        \hline
        \hline
        SINDy (Second Order) & 76 (\textbf{94})\\
        \hline
        PWA (\mycite{FERRARITRECATE2003205}) & 75\\
        \hline
        Jump (\mycite{Bemporad2018}) & 83\\
        \hline
	\end{tabular}
\end{table}

\section{Hidden, non-smooth dynamics: the Bouc-Wen hysteresis model}\label{sec:bouc-wen} 
We now consider the dataset described in~\cite{Schoukens2017,2020noelHystereticBenchmarkDynamic}, which consists of simulated data from the model proposed in~\cite{Bouc1967ForcedVO}. The former describes a nonlinear oscillator with dynamic memory, described by the system
\begin{subequations}\label{eq:bw_model}
    \begin{align}
        & m_L \ddot{y}(t) = u(t) - c_L \dot{y}(t) - k_L y(t) - z(t), \label{eq:bw-observed-state}\\
        &\dot{z}(t) = \alpha \dot{y}(t) - \beta (\gamma |\dot{y}(t)| z(t) + \delta \dot{y}(t) |z(t)|), \label{eq:bw-hidden-state}
    \end{align}
\end{subequations}
where $y(t)$ [m] is the output displacement, and the control input $u(t)$ [N] is an external force. The system thus features a hysteretic behavior, whose characteristics depend on the (unknown) parameters\footnote{Their true value can be found in \cite{2020noelHystereticBenchmarkDynamic}.} $m_L,\ c_L,\ k_L,\ \alpha,\ \beta,\ \gamma,\ \delta$.\vspace{-.5cm}

\paragraph*{\textbf{Challenges and Goal.}} As indicated in \cite{2020noelHystereticBenchmarkDynamic}, the main challenges of this benchmark example are: $(i)$ the dependence of the nonlinearity on a non-measurable internal variable $z(t)$, $(ii)$ the fact that the nonlinearity is dynamic, i.e., it has its own memory, and finally $(iii)$ that~\eqref{eq:bw-hidden-state} does not admit finite Taylor series expansion. Challenges $(i)-(ii)$ defy the foundational assumption of the standard SINDy method, since regression cannot be performed against a non-measured variable.
Challenge $(iii)$ is even more insidious as no finite basis expansion of polynomials can suitably recover~\eqref{eq:bw-hidden-state}, meaning the choice of candidate terms becomes more involved. Given these features, our goal is to show how these limitations can be overcome by leveraging the structure of the mathematical model in \eqref{eq:bw_model} within the framework of the SINDy method.\vspace{-.5cm}

\paragraph*{\textbf{Hands-on strategy.}} A \emph{naïf} approach, neglecting the unobserved state would be applicable, leads to:
\begin{equation}
    \ddot{y}(t) = \phi^{BW}(\dot{y}(t), y(t), u(t))^{\top}\theta,
\end{equation}
such a model is doomed to failure. Indeed, in quasi-static forcing conditions, the restoring force given by this model would only depend on displacement-input pairs, thus not exhibiting the hysteretic behavior.\\
As a possible workaround to this limitation, we have turned the training procedure into an iterative process, exploiting the structure of equation~\eqref{eq:bw-observed-state}. Specifically, we
\begin{enumerate}
    \item guess the values for $m_L$, $k_L$ and $c_L$, thus fully characterizing \eqref{eq:bw-observed-state} (that is hence fixed);
    \item learn the equation for the unobserved state $z(t)$ with SINDy, by setting $$z(t) = u(t) - c_L \dot{y}(t) - k_L y(t) - m_L \ddot{y}(t),$$ and learning a model of the form
    \begin{equation}
        \dot{z}(t)=\phi^{z}(z(t),\dot{y}(t))^{\top}\theta_z,
    \end{equation}
    using a library $\phi^{z}$ comprising products of $\dot{y}(t)$, $z(t)$, $|\dot{y}(t)|$, and $|z(t)|$;
    \item accordingly simulate the full system.
\end{enumerate}
This procedure is iterated for different guesses of $m_L$, $k_L$, $c_L$, selecting their \textquotedblleft best\textquotedblright \ values as the one minimizing the RMSE in validation. 
We stress that 
relying more on our (partial) prior knowledge of the system behavior is aligned with strategies already proposed to tackle unobserved variables (see, e.g.,~\mycite{2020reinboldUsingNoisyIncomplete}).\\vspace{-.5cm}

\begin{figure}[tb!]
    \centering
    \includegraphics[scale=.45]{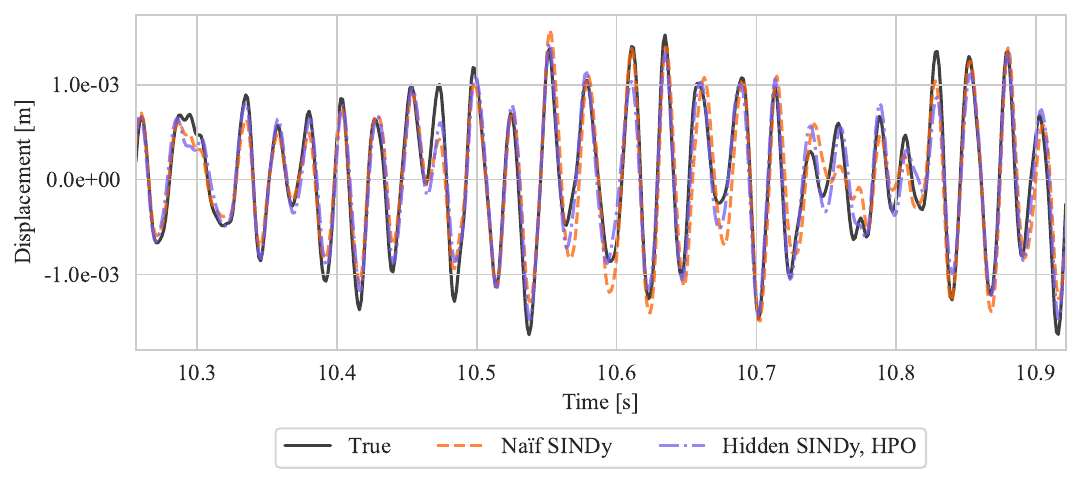}
    \caption{Bouc-Wen: simulation snapshot with multi-sine $u$.}\vspace{-.2cm}
    \label{fig:bw-validation}
\end{figure}

\paragraph*{\textbf{Results.}} The use of Bayesian optimization to search for the best guesses\footnote{The search space is limited to positive parameters.} of $m_{L}$, $k_{L}$  and $c_{L}$ yields\vspace{.1cm}
\begin{subequations}
\begin{align*}
    \nonumber &\dot{z}(t)\!=\! \alpha\ \dot{y}(t) - \beta\ z(t)|\dot{y}(t)| + \gamma\ |z(t)|\dot{y}(t) - \delta\ |z(t)| |\dot{y}(t)|,\vspace{.1cm}\\
    & \ddot{y}(t)\!=\! -0.213 \dot{y}(t)\! -\! 27685.724 y(t)\!-\! 0.379 z(t) \!+\! 0.379 u(t),\vspace{.1cm}
\end{align*}
\end{subequations}
with $\alpha = 68815.219, \beta = 570.725, \gamma = 700.135, \delta = 0.815$.
By comparing this model (referred to as \emph{Hidden SINDy, HPO}) with that attained with the \emph{na\"ive} approach in testing (see~\figurename{~\ref{fig:bw-validation}} and \tablename{~\ref{tab:bw-results}}), we see that the former does not lead to excessively poor simulation performances. Nonetheless, the obtained model is qualitatively wrong. To see this, we create an additional test trajectory by forcing the system with a low-frequency ($f = 0.75$ Hz) sinusoidal input of amplitude $A = 150\ N$ for the duration of one period ($T = 1.\bar{3}\ s$).
The resulting displacement-force plot can be seen in \figurename{~\ref{fig:bw-hysteresis}}, wherein the \emph{na\"ive} SINDy approach fails in replicating the hysteretic behavior, as expected.
Further, we consider the model retrieved in an \emph{ideal} scenario (
denoted as \emph{Hidden SINDy, Ideal}), where $m_{L}$, $k_{L}$ and $c_{L}$ are set to their true values, namely\vspace{.1cm}
\begin{subequations}
\begin{align*}
    & \dot{z}(t)\! =\! 59835.845 \dot{y}(t) \!-\! 442.497 z(t)|\dot{y}(t)| \!+\! 357.725 |z(t)| \dot{y}(t),\vspace{.1cm}\\
    & \ddot{y}(t) = - 5\dot{y}(t) - 25000 y(t) - 0.5 z(t) + 0.5 u(t).\vspace{.1cm}
\end{align*}
\end{subequations}
It can be noticed that only in this last case the correct structure of equation~\eqref{eq:bw-hidden-state} is discovered (while \emph{Hidden SINDy, HPO} features the additional term $|z(t)||x(t)|$). Hence, the hysteresis cycle reconstructed in \figurename{~\ref{fig:bw-hysteresis}} more closely resembles the actual one. However, as highlighted by the metrics reported in \tablename{~\ref{tab:bw-results}}, this model surprisingly performs worse than the \emph{Hidden SINDy, HPO} model in testing, especially 
for multi-sine and sine sweep inputs.\vspace{-.5cm}

\paragraph*{\textbf{Take-home message.}}
Even if the \textit{Hidden SINDy, HPO} model is the best-performing one in both validation and the test sets, it still cannot capture the true hysteretic behavior properly. This is possibly due to the ill-conditioning of the regression problem involving $\dot{z}(t)$ stemming from the different scales of the parameters and variables, as well as to the potential collinearity between $z(t)|\dot{y}(t)|$ and $|z(t)|\dot{y}(t)$. Therefore, despite the considerable priors exploited throughout learning, approaches based on derivative fitting eventually fail. At the same time, we strikingly obtain that the best model according to our indicators and the reconstructed hysteretic cycle is the one that does not match the structure of the true system. To truly discover the structure of the underlying system we require even stronger priors (i.e., the exact knowledge of part of the parameters and dynamics) leading nonetheless to poorer performance.
\arnew{
\begin{figure}[tb!]
    \centering
    \includegraphics[scale=.45]{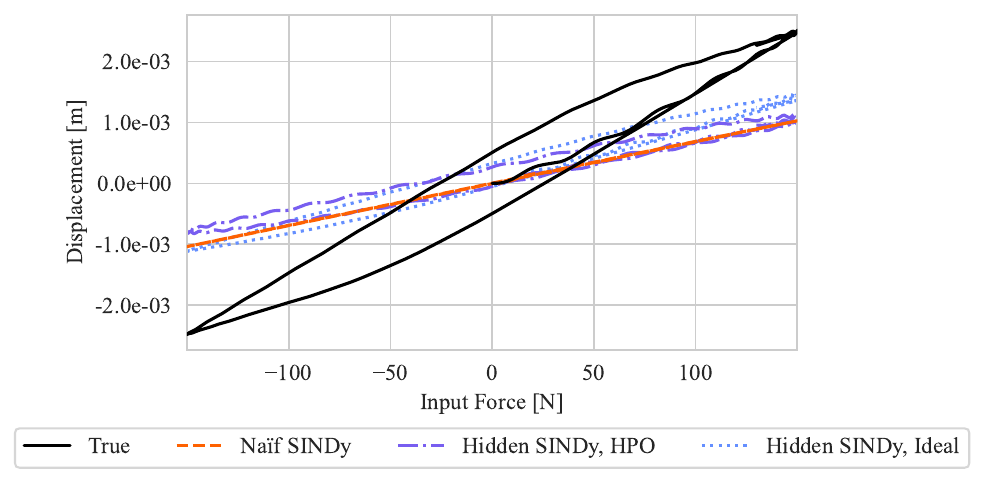}
    \caption{Bouc-Wen model: Hysteresis cycles.}
    \label{fig:bw-hysteresis}
\end{figure}}
\begin{table*}[!tb]
\caption{Bouc-Wen model: RMSE ($\times 10^{-4}$)~[m] on the test sets.}\label{tab:bw-results}
    \centering
    \begin{tabular}{lccc}
      \textbf{Model}
     & \textbf{Low-frequency test} & \textbf{Multi-sine test} & \textbf{Sine Sweep test} \\
     \hline
     \hline
    Na\"if SINDy & 10.99 & 2.75         &  3.32         \\
    \hline
    Hidden SINDy, HPO (Ideal)       &  11.26 (\textbf{8.87}) & \textbf{1.83} (2.52) & \textbf{1.72} (3.03)         \\
    \hline
    \end{tabular}
\end{table*}

\section{Control inputs acting on unobserved states: the Cascaded Tanks example}\label{sec:cascaded-tanks} 
Our final tests are carried out on the benchmark dataset described in \mycite{Schoukens2017}; \mycite{cascadedTanksBenchmark}, comprising input/output data collected from a cascade of two tanks, interconnected via a valve. The upper tank is fed from a water reservoir through a pump, whose input voltage $u(t)$ [V] represents that of the overall system. The water level of the lower tank is measured via capacitive water level sensors, making its voltage $y(t)$ [V] the measured output. Note that the tanks are both open at the top, leading to non-smooth dynamics of the overall system in the event of overflows.\\
By relying on Bernoulli’s principle, the system's behavior \emph{far from saturation} well-approximated by the following mathematical model:
\begin{subequations}\label{eq:tanks-hf}
\begin{align}
    & \dot{x}_{1}(t)=-k_{1}\sqrt{x_1(t)}+k_2u(t),\label{eq:ctanks-hidden-state}\\
    & \dot{x}_{2}(t)=k_3\sqrt{x_1(t)}-k_{4}\sqrt{x_2}(t),\label{eq:ctanks-observed-state}\\
    &y(t)=x_{2}(t),\label{eq:ctanks-observed-state-2}
\end{align}
\end{subequations}
where $x_{i}(t)$ is the water level in the $i$-th tank, for $i=1,2$, while $\{k_{i}\}_{i=1}^{4}$ are unknown parameters. This model, however, is no longer valid when saturations intervene, calling for strategies that surpass a pure gray-box approach.
\vspace{-.5cm}

\paragraph*{\textbf{Challenges and goal.}}
As discussed in~\cite{cascadedTanksBenchmark}, this benchmark case study involves $(i)$ a combination of hard saturations with the weakly nonlinear behavior of the system, and $(ii)$ a possible overflow from the upper to the lower tank, in turn representing an input-dependent process noise. On top of these issues, $(iii)$ one of the states (i.e., the level of water in the upper tank) is unknown and, thus, it has to be guessed, and $(iv)$ the data record available for learning is short.   
The first three issues are especially challenging in applying the SINDy method as saturations are particularly difficult to model with SINDy, unless embedded in the candidate library. At the same time, bases crafted to \emph{encode} discontinuities might lead to difficulties in the fitting process. Moreover, the possible presence of outflows and the lack of information on the upper tank entail that the state of the system is not fully measurable. Notably, the control input acts directly on the unmeasured state, governing the inflow to the lower tank. This is a critical issue because no library that is restricted to the observed state and input can recover a qualitatively correct input/output dependency.
Contrarily to the Pick and Place example, the unobserved state cannot be trivially reconstructed from the observed one. Our goal is thus to check how varying levels of physical priors (based on equations~\eqref{eq:tanks-hf}) can help to cope with these limitations.\vspace{-.5cm}

\paragraph*{\textbf{Hands-on strategy.}}
To overcome the limitations of the simplistic input/output model of the form
\begin{equation}\label{eq:naif_tank}
	\dot{y}(t)=\phi^{\mathrm{tanks}}(y(t),u(t))^{\top}\theta,
\end{equation}
in the presence of hidden states and unknown initial conditions, we tap into a supposed deeper knowledge of the hidden dynamics, i.e.,
we assume to know the structure of~\eqref{eq:ctanks-hidden-state} and the saturation points for $x_1$.
Hence, we articulate the fitting procedure into finding the values of the parameters $k_1$ and $k_2$ characterizing the hidden dynamics, as well as an expression of equation~\eqref{eq:ctanks-observed-state} through SINDy. This is performed by: \vspace{-.2cm}
\begin{enumerate}
    \item guessing $k_1$, $k_2$ and $x_1(0)$;
    \item simulating the unobserved state as $\hat{x}_1(t)$;
    \item fitting a SINDy model 
    \begin{equation}
    \dot{y}(t)=\phi^{\mathrm{tanks}}(y(t), \hat{x}_1(t), u(t))^{\top}\theta,
    \end{equation}
    for the observed state $y(t)$.
\end{enumerate}
This procedure iterates over different guesses of $k_1$, $k_2$, and induced $\theta$, selecting their best combination based on the RMSE in validation. This involves estimating $x_1(0)$ over validation (and test) trajectories. To do so, we augment the initial condition with the derivative of the observed state $\dot{y}(0)$, selecting the estimate $\hat{x}_1(0)$ as
\begin{equation}
    \hat{x}_{1}(0) \leftarrow \arg\!\min_{x_{1}}~(\dot{y}(0) - \phi^{\mathrm{tanks}}(y(0), x_1, u(0))^{\top}\theta)^2.
\end{equation}
This optimization problem can be eventually replaced by a simple grid search within the physical bounds for ${x}_{1}(0)$.\vspace{-.5cm}

\begin{figure}[!tb]
	\centering
	\includegraphics[scale=.45]{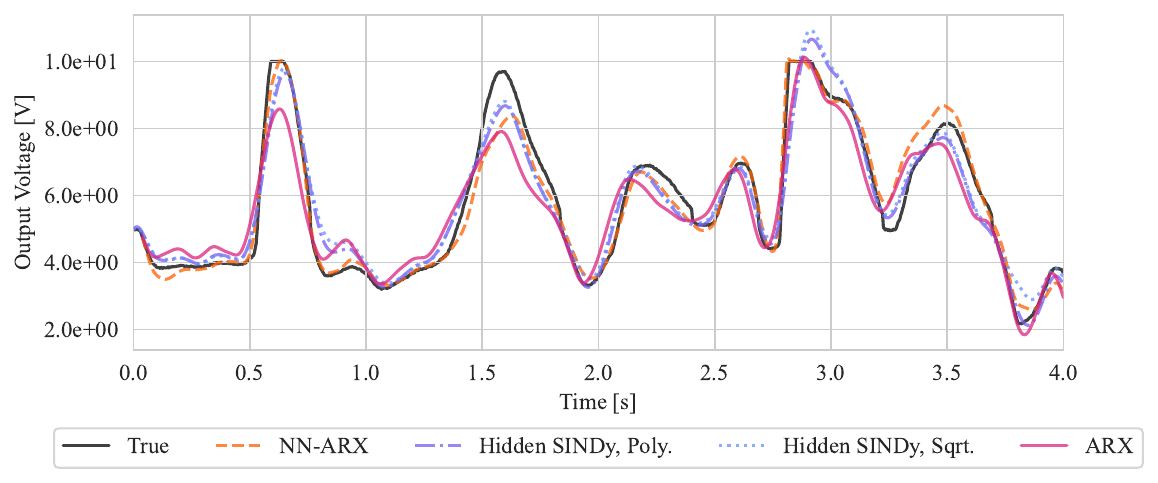}\vspace{-.3cm}
	\caption{Cascaded tanks: simulations over the test set.}\label{fig:tanks}
\end{figure}

\paragraph*{\textbf{Results.}}
\begin{table}[!tb]
\caption{Cascaded tanks: performance metrics over the test set.}\label{tab:tanks}
	\centering
	\begin{tabular}{l|cc}
 \textbf{Model} & \textbf{BFR [\%]} & \textbf{RMSE} \\
 \hline
 \hline
    SINDy, Poly. (Sqrt.)        & 73.16 (53.09)                                                           & 1.09 (1.44)                            \\
    \hline
    SINDy Second Order &  0 & 2.12 \\
    \hline
    Hidden SINDy, Poly. (Sqrt.)      & \textbf{93.24} (92.82)                                                           & \textbf{0.55} (0.56)                             \\
    \hline
    NN-ARX (ARX)                & \textbf{95.10} (86.55)                                                           & \textbf{0.47} (0.77)\\
    \hline
\end{tabular}
\end{table}
As a first attempt to understand the level of prior needed to obtain satisfactory simulation performance, we stick to the \emph{na\"ive} input/output model in \eqref{eq:naif_tank} comparing the performance attained with a fully uninformed candidate library choices with those inspired by~\eqref{eq:tanks-hf}. To this end, we include the square root of the input and the output in the library, together with more conventional second-order polynomial functions. Meanwhile, to prevent numerical issues during simulation, we threshold the output before computing the square root based on a range that we assume is known.
By using training and validation sets comprising $768$ and $256$ input/output samples\footnote{The validation set is obtained by extracting the last $256$ samples from the original training data from~\cite{cascadedTanksBenchmark}.}, respectively, the model recovered by using polynomial functions only (denoted as \emph{SINDy, Poly.}) is
\begin{align}\label{eq:ctanks-sindy-naive}
    \dot{y}(t) &= -4.822 y(t) + 0.740 y(t) u(t) + 1.742 u^2(t),
\end{align}
whereas, including the square roots, we get
\begin{subequations}\label{eq:ctanks-sindy-sqrt}
    \begin{align}
        &\dot{y}(t) = [u(t), \sqrt{y(t)}, y(t) u(t), u^2(t)] \cdot \theta_{\mathrm{SINDy,\ Sqrt.}}\\
        &\theta_{\mathrm{SINDy,\ Sqrt.}} = [-8.594 , -1.831, 0.426, 3.999]^\top
    \end{align}
\end{subequations}
showing that SINDy is indeed capable of discovering the nonlinearity of the observed output when the set of bases is properly chosen. Despite this, when assessing the quality of the previous models on the test set, the one incorporating the stronger priors (labeled as \emph{SINDy, Sqrt.}) is worse in reproducing the system behavior, although the performance of polynomial SINDy cannot be deemed satisfactory either (see \tablename{~\ref{tab:tanks}}). These results are not surprising, considering that these models lack the expressive power needed to model the system in saturation. 
We thus employ the presented hands-on solution to overcome this limitation, by still considering the two sets of basis functions used for the na\"ive input/output models to describe the output dynamics. Using the polynomial bases, we obtain
\begin{subequations}\label{eq:ctanks-hidden-naive}
    \begin{align}
        &\dot{x}_1(t) = -43.290 \sqrt{x_1(t)} + 33.353 u(t),\\
        &\dot{y}(t) = -10.100 y(t) + 9.330 x_1(t) + 0.421 x_1^2(t),
    \end{align}
\end{subequations}
while the model obtained considering the square roots as candidates is given by
\begin{subequations}\label{eq:ctanks-hidden-sqrt}
    \begin{align}
        &\dot{x}_1(t) = -48.204 \sqrt{x_1(t)} + 36.012 u(t),\\
        & \dot{y}(t) = [y(t), \sqrt{y(t)}, x_1(t), \sqrt{x_1(t)}]\cdot\theta_{\mathrm{H. SINDy,\ Sqrt.}}\\
        &\theta_{\mathrm{H. SINDy,\ Sqrt.}} = [-17.904, 42.235, 25.654, 54.938]^\top
    \end{align}
\end{subequations}
Once again this shows that SINDy can discover the nonlinearity featured in the dynamics of the observed state\footnote{This is \textquotedblleft partially\textquotedblright \ true, as the model with polynomial bases only results in an RMSE in validation of $0.356$, against the $0.383$ attained by considering the square root basis, ultimately leading to the choice of the former (\textquotedblleft wrong\textquotedblright) model.}. Differently from the na\"ive case, the approach yields comparable performance for the two choices of libraries (\emph{Hidden SINDy, Poly.} and \emph{Hidden SINDy, Sqrt.} respectively), with the former being slightly superior as shown by both \figurename{~\ref{fig:tanks}} and \tablename{~\ref{tab:tanks}}.
To conclude our analysis, we now compare the performance of the models retrieved with SINDy with those obtained by an autoregressive model with exogenous input (ARX) whose order is chosen via Bayesian optimization, and a Neural Network AutoRegressive model with eXternal input (NN-ARX), similarly to what was performed in \cite{2018wordenEvolutionarySystemIdentification}. Although the choice of the (linear) ARX model is far from optimal in this application, this simple model still outperforms the nonlinear ones identified using SINDy in the input/output approach (i.e., models~\eqref{eq:ctanks-sindy-naive},~\eqref{eq:ctanks-sindy-sqrt}), as clearly shown in~\tablename{~\ref{tab:tanks}}, further spotlighting the limitations of the standard SINDy method in the presence of unobserved states. Meanwhile, the NN-ARX model performs slightly better than the best model obtained with our hands-on strategy, at the price of reduced interpretability and (based on our experience) a higher sensitivity to hyper-parameters. \vspace{-.25cm}     

\paragraph*{\textbf{Take-home message.}} The results show that, much like the previous case study, without a deep insight into the system's structure, SINDy fails to simulate the dynamic behavior accurately. Furthermore, model selection based on simulation error minimization leans towards a qualitatively incorrect choice of candidate terms, leaving out those chosen based on background knowledge of the system. Hence, a careful choice of bases is crucial, yet possibly not sufficient, for the approach to unveil the underlying dynamics of the system in the presence of hidden states.

\section{Final remarks \& promising directions}\label{sec:open-challenges}
After a brief review of the main theoretical aspects of SINDy, in this work we have focused on analyzing the performance of this algorithm on three benchmark datasets for nonlinear system identification. Although SINDy has several appealing properties, e.g., the capability of returning sparse models, that can be easily interpreted and analyzed, our tests have spotlighted the deterioration of model accuracy in the absence of full state observability, the potential difficulty in crafting appropriate libraries, especially to characterize hard nonlinearities. Indeed, as exemplified by the Bouc-Wen case, using derivative-fitting methods complicates learning and sets practical limits to the choices of library terms. To provide quick fixes to these limitations, for each case study we have thus indicated practical approaches, often (excessively) reliant on partial knowledge of the dynamics, but leading to better models (qualitatively, quantitatively, or both).\\ 
Our tests highlight that the full-observability requirement is perhaps one of the most limiting and potentially critical for the method, for which there is no definitive solution in literature, especially if one aims to preserve the meaning of the original problem variables. As a consequence, two of the most promising research opportunities consists of equipping SINDy with state estimation capabilities or devising ways to introduce extended states in the SINDy formulation (e.g., as done for Neural Ordinary Differential Equations by \mycite{2022rahmanNeuralOrdinaryDifferential}), potentially forsaking the efficiency of directly fitting a model on the derivatives to guarantee more powerful and insightful results.\\ 
To conclude, the SINDy method and its extensions provide interesting building blocks for complex model learning, leading to interpretable models and involving fewer hyper-parameters than most alternatives. However, expert oversight and modeling experience are still very much required to attain performances comparable to other black-box approaches.


\begin{ack}\vspace{-.2cm}
We thank Anastasia Bizyaeva and Davide Fleres for their insights and perspectives on the SINDy method. The work was partially supported by the project 4D Drone Swarms
under grant no. F/310097/01-04/X56,  PRIN project TECHIE, Cod. 2022KPHA24 CUP: D53D23001320006 and by the Sustainable Mobility National Research Center, Spoke 5: “Light Vehicles and Active Mobility”.
\end{ack}

\bibliography{Benchmarking_SINDy_ArXiv}             
                                                   








\end{document}